# Crystal structure of the unconventional spin-triplet superconductor UTe$_2$ at low temperature by single crystal neutron diffraction.

V. Hutanu[1], H. Deng[1], S. Ran[2,3], W. T. Fuhrman[3], H. Thoma[4], N. P. Butch[2,3]

[1] Institute of Crystallography, RWTH Aachen University and Jülich Centre for Neutron Science (JCNS) at Heinz Maier-Leibnitz Zentrum (MLZ), D-85748 Garching, Germany

[2] NIST Center for Neutron Research, National Institute of Standards and Technology, Gaithersburg, MD 20899, USA

[3] Center for Nanophysics and Advanced Materials, Department of Physics, University of Maryland, College Park, MD 20742, USA

[4] Jülich Centre for Neutron Science (JCNS) at Heinz Maier-Leibnitz Zentrum (MLZ), D-85748 Garching, Germany

The crystal structure of the new superconductor UTe$_2$ has been investigated for the first time at low temperature (LT) of 2.7 K, just closely above the superconducting transition temperature of about 1.7 K by single crystal neutron diffraction, in order to prove, whether the orthorhombic structure of type *Immm* (Nr. 71 Int. Tabl.) reported for room temperature (RT) persists down to the superconducting phase and can be considered as a parent symmetry for the development of spin triplet superconductivity. Our results show that the RT structure reported previously obtained by single crystal X-Ray diffraction indeed describes also the LT neutron diffraction data with high precision. No structural change from RT down to 2.7 K is observed. Detailed structural parameters for UTe$_2$ at LT are reported.

**Introduction**

The crystal structure of the actinide compound UTe$_2$ was investigated for the first time by Klein et al. in the early seventies [1]. Using X-ray powder diffraction they determined the orthorhombic space group (SG) *Immm* and the lattice parameters. Later using single crystal X-ray diffraction Ellert at al. [2] confirmed the orthorhombic type of the structure, but concluded on a different SG, namely *Pnnn*. The first exhaustive crystallographic study using single crystal x-ray diffraction was done in 1988 [3]. The authors reconfirmed the SG *Immm* and reported precise crystallographic details like lattice parameters, fractional atomic coordinates, anisotropic atomic displacement parameters as well selected bond length, determined at room temperature (RT). In the early 2000's, motivated by interest in the magnetic properties of the uranium compounds, the strongly anisotropic structure from Ref. [3] was reinvestigated and confirmed, using the same experimental method and providing very similar structural results at RT [4]. Very recently unconventional spin-triplet superconductivity has been reported in UTe$_2$ below 1.6 K [5]. It was proposed that it belongs to the family of U based unconventional ferromagnetic superconductors as a paramagnetic end member of this series, where spin fluctuations without an ordered magnetic moment, play a major role in Cooper pairing. In this regard the question occurs, whether the reported RT crystal structure continues to persist down to the very low temperature of the superconducting transition and in how far the structural parameters change under those conditions. To our best knowledge, there are no reports about thermal evolution of the crystal structure in UTe$_2$ so far. Moreover, detailed structural studies on this compound were never performed using neutron diffraction. The neutron scattering happens mostly from the nuclei, while X-rays scatter from the electron shells. Thus, the latter one is more suitable for determining the charge distribution or charge ordering, while the first one is superior in determining the atomic positions and displacement parameters. Also, because of their higher penetration ability, neutrons are commonly used at very LT, where sample is placed in a cryostat with metal walls. In order to provide structural details on UTe$_2$ at LT close to the superconducting phase transition, we performed a single-crystal neutron diffraction experiment, using short wave length neutrons.

**Experimental**

High quality single crystals of UTe$_2$ with triangular plate-like shape and typical size up to 3x3x1 mm and mass between 20-60 mg were obtained by the chemical vapour transport method. A crystal from the same growth badge as those described in the Ref. [5] was used for the present study. Single-crystal neutron diffraction was performed on the single crystal diffractometer POLI [6] at the hot neutron source of the FRM II reactor at the Heinz-Maier-Leibnitz Zentrum (MLZ) Germany. A short wavelength of 0.9 Å from a Cu (220) monochromator was used in order to reduce potential parasitic effects of absorption and extinction. A $^3$He point detector, optimised for neutrons with short wavelengths was used for these measurements. The sample was wrapped in Al foil to ensure homogeneity of its temperature and placed in a liquid He variable temperature insert cryostat (Oxford Instruments). The temperature was measured and controlled by a diode sensor near the sample position and a stability of better 0.1 K was achieved. The corrected integrated intensities of the measured reflections were obtained with the DAVINCI program [7]. Refinement of the structural parameters were performed with the program JANA2006 [8]. Using a large new cryomagnet at zero field as a cryostat and lifting mechanics for single tube detector in combination with short neutron wavelength on the diffractometer POLI [6], the reasonably large angular access to the reciprocal space is provided. The sample was cooled down to 2.7 K and centred in the vertical position. In a preliminary quick test-scan a total number of 448 Bragg reflections with sin $\theta/\lambda \leq 0.63$ Å$^{-1}$ were collected. As a result of the test 327 reflections satisfying the criterion $I_{max} > 1.75\ I_{bgr.}$ were selected for further detailed measurement. The selected from the test scan peaks were individually pre-centered and carefully measured by omega scan. After visual inspection and profile analysis of the measured peaks the total number of 298 proper-centred Bragg reflections satisfying the criterion I > 10$\sigma$(I) were used for the refinement. Experimental and refinement details are summarized in Table 1.

**Results**

The number of special peaks of type h00, 0k0, hk0, h0l, satisfying the condition h+k+l =2n+1, violating I-centring were carefully scanned both at 2.7 K and 1.7 K (lowest temperature of the cryostat). No evidence of those peaks and correspondingly no such violation of extinction rules was observed. Starting parameters for the least-squares refinement were obtained from the RT structure determined previously in the orthorhombic SG *Immm* by x-ray single-crystal diffraction [3]. The quality of the fit, assuming SG *Immm* also at LT is shown in the Fig. 1. As-measured and not symmetry-averaged data are shown. Using averaging of symmetry equivalent peaks 133 independent reflections result. Structure refinement on the averaged data leads to even lower $R_F$=0.013. The high fit quality for our LT neutron data using RT model (with adjusted parameters), may be linked to careful data collection on the one side and perfectly matching structural model on the other. Table 2 presents the refined atomic coordinates, and both the isotropic and anisotropic atomic displacement ($U_{iso}$, $U_{ij}$) parameters correspondingly. Full details of the refinement, including bond lengths and angles, are provided in the deposited crystallographic information file (CIF) [9].

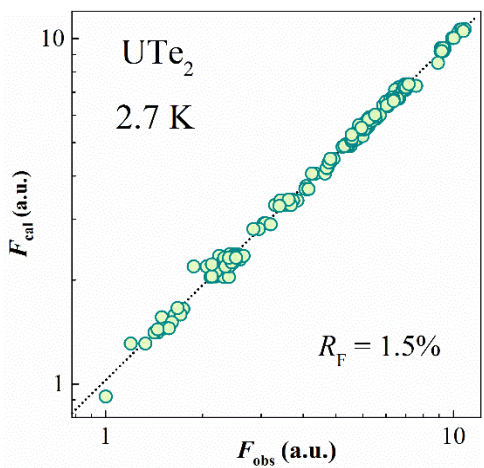

Figure 1. (Color online) Quality of the diffraction data refinement for the nuclear structure of UTe$_2$ at 2.7 K in SG *Immm*. The experimental integrated intensities ($F_{obs}$) are plotted against the calculated ones ($F_{calc}$) in log scale for better visualisation of the weak reflections.

Table 1. Single-crystal neutron diffraction experimental and refinement details

| Crystal data | |
|---|---|
| chemical formula | $UTe_2$ |
| relative molar mass | 493.23 |
| cell setting, space group | orthorhombic, Immm |
| T (K) | 2.7 |
| a, b, c (Å) | 4.123, 6.086, 13.812 |
| V (Å$^3$) | 346.579 |
| Z | 4 |
| Dx (Mg m$^{-3}$) | 9.1993 |
| μ (mm$^{-1}$) | 0.0095 |
| crystal form, color | plate-like, black |
| crystal size (mm) | 3 x 3 x 1 |
| Data collection | |
| diffractometer | normal-beam diffractometer POLI |
| radiation source | nuclear reactor |
| monochromator | Cu(220) |
| radiation type | constant wavelength neutron, |
| wavelength (Å) | 0.904(1) |
| data collection method | ω - scans |
| [sin θ/λ]$_{max}$ (Å$^{-1}$) | 0.63 |
| range of h, k, l | -5 → h→5 |
| | -6 → k→7 |
| | -9 → l→13 |
| no. of measured reflections | 327 |
| no. of observed reflections with I >10σ(I) | 298 |
| no. of independent reflections with I > 10σ(I) | 133 |
| $R_{int}$ | 1.86 |
| Refinement | |
| refinement on | F |
| R[F > 3σ(F)], wR(F), S | 0.015, 0.021, 1.52 |
| no. of parameters | 14 |
| weighting scheme, w | 1/[σ2(F) + 0.0001F2] |
| extinction correction | Isotropic, Gaussian type 1 |
| extinction coefficient | 0.037(2) |

Table 2. Fractional atomic coordinates (x, y, z), isotropic and anisotropic atomic displacement parameters (U in Å$^2$) for $UTe_2$ at 2.7 K refined in the orthorhombic SG *Immm* according to the present single-crystal neutron diffraction data. In this model $U_{12} = U_{13} = U_{23} = 0$.

| Atom | W.p | x | Y | z | $U_{11}$ | $U_{22}$ | $U_{33}$ | $U_{iso}$ |
|---|---|---|---|---|---|---|---|---|
| U | 4i | 0.00000 | 0.00000 | 0.13473(6) | 0.0021(2) | 0.0019(3) | 0.0014(5) | 0.0018(2) |
| Te(1) | 4j | 0.50000 | 0.00000 | 0.29799(10) | 0.0033(3) | 0.0035(4) | 0.0034(8) | 0.0033(3) |
| Te(2) | 4h | 0.00000 | 0.25062(13) | 0.50000 | 0.0035(3) | 0.0039(4) | 0.0031(8) | 0.0035(3) |

Fig. 2 shows the perspective view of the $UTe_2$ crystal structure. The positions of the atoms are shown by the ellipsoids of the refined anisotropic displacement parameters (ADP) with probability as high as

99%. Small, almost spherical displacement parameters, showing no significant elongations, are observed for Te atoms independent of the Wyckoff position. Even smaller parameters are refined for U atoms. No strong anisotropy in the thermal displacement parameter in UTe$_2$ structure was reported also for the RT [3]. However, the results regarding those parameters obtained by x-ray and neutron diffraction have different physical meaning, taking into account the different nature of their respective scattering. A big advantage of neutron diffraction is the lack of form factor-decay and (almost) no absorption. Reported here ADP are one order of magnitude more precise.

Comparing the atomic positions in the unit cell at RT and LT respectively, one observes that fractional coordinates for Te from our data are precisely reproducing (within one sigma error bars) those reported at RT by Refs. [3,4]. Which are also very similar to each other for both RT X-ray measurements. For z-coordinate of the U, however, small discrepancy (at the level of 4 x sigma) between reported RT X-ray results [3,4] exists. The z-coordinate of the U position in our results at LT is smaller (at the level of 10 x sigma) in comparison to the calculated mean value for RT from Ref.[3,4]. The first coordination-sphere polyhedron of the U atom, which is interesting to consider in terms of the origin of the superconductivity and/or potential magnetic interactions, consists of four same-length U-Te(2) bonds and two shorter U-Te(1)s and two longer U-Te(1)l bonds correspondingly, as shown in Fig. 3. The main difference between RT and LT structure provoked by the shift of the U position results in the relative stretching of the U-Te(1)l bond: U-Te(1)l/U-Te(2) = 1.0053(2) at LT comparing to 1.0038(4) at RT. Meanwhile, the ratio between U-Te(1)s and U-Te(2) does not change: U-Te(2)/U-Te(1)s = 1.0360(5) at RT versus 1.0358(5) at LT. Also, other bond-length ratios do not changes significantly.

The lattice parameters were obtained by refinement of the orientation matrix using angular positions of the strongest 200 centred Bragg reflections and fixed known off-sets for the instrument axes. A comparison between the results for the lattice parameters at RT and LT are shown in Table 3. The lattice shrinks upon cooling, as seen from the refined lattice parameters, relatively homogeneously (within error of the measurement) in all directions: $\Delta a/a = 0.87(13)\%$, $\Delta b/b = 0.62(16)\%$, $\Delta c/c = 0.96(14)\%$. This corresponds to about 2.5% volume reduction and approximated maximal linear coefficient of thermal expansion of about $\alpha = 2.8(7)\ 10^{-5}\ K^{-1}$.

Table 3. Crystal lattice parameters of UTe$_2$ at room and low temperature respectively.

| a / Å | b / Å | c / Å | Temperature / K | Determined by | Source |
|---|---|---|---|---|---|
| 4.123(5) | 6.086(9) | 13.812(17) | 2.7 | Single crystal, neutrons | this study |
| 4.159(1) | 6.124(2) | 13.945(9) | 295 | Single crystal, X-rays | [3] |

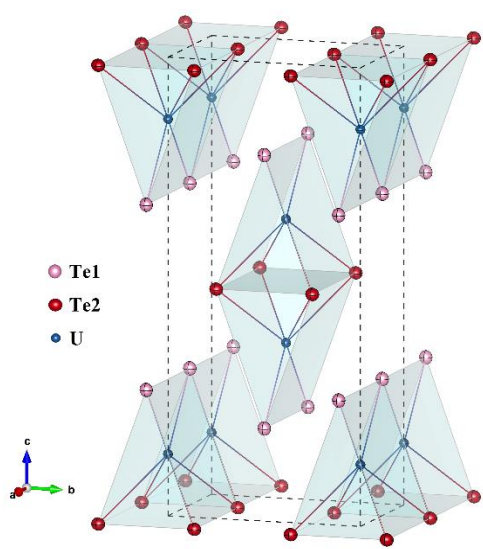

Figure 2. Perspective view of the crystal structure of UTe$_2$ in SG *Immm*. VESTA software [11] was used for visualization.

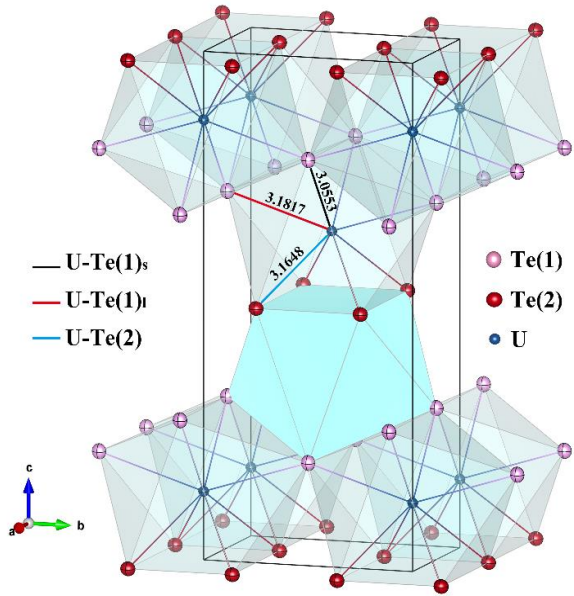

Figure 3. The first coordination-sphere polyhedron of U (cation) by neighbouring Te (anions) in UTe$_2$ with bond length at 2.7 K. The noted bond length are in Å.

## Discussion

Our single crystal neutron diffraction results are consistent with previously measured electrical resistivity, magnetization, and specific heat data over a wide temperature range [5]. All evidence points to the absence of both structural and magnetic phase transitions in UTe$_2$ between room temperature and 2.7 K. Instead, the large temperature dependence of the transport properties and the magnetic anisotropy are the result of strongly interacting uranium-based f-states. This fact is reflected in the observed relatively large linear thermal expansion coefficient and pronounced change of the c lattice-parameter, together with the change in the U-position z-coordinate. Crucially, there is no magnetic order in UTe$_2$ in the normal state, which makes this superconductor qualitatively different from ferromagnetic URhGe, UCoGe, and UGe$_2$ [10] despite the similar anisotropy in superconducting upper critical fields and superficial crystal structure similarities. Our new diffraction data also support the picture of UTe$_2$ as a quantum critical ferromagnet, as there is no evidence for antiferromagnetic order that could produce the unusual field-temperature scaling of the magnetic susceptibility reported earlier [5]. The novel emergence of spin-triplet superconductivity from a paramagnetic normal state characterized by strong ferromagnetic spin fluctuations calls for focused theoretical attention.

## Conclusion

Our neutron diffraction results prove that no structural change occurs in UTe$_2$ at low temperatures between RT down to 2.7 K, close to the superconducting transition temperature. An orthorhombic structural model with SG *Immm* (Nr. 71 Int. Tabl.) well describes the observed data at LT. Detailed structural parameters for UTe$_2$ at LT are reported for the first time and provide fundamental input for the further experimental investigations and theoretical modelling on this interesting superconducting material.

## Acknowledgements

This work is based on experiments performed at instrument POLI at Heinz Maier-Leibnitz Zentrum (MLZ), Garching, Germany, operated by RWTH Aachen University in cooperation with FZ Jülich (Jülich-Aachen Research Alliance JARA). The main author is grateful to NIST Center for Neutron Research for the hospitality he enjoyed during his stay on the Exchange Visitor Program in Gaithersburg, where the idea and plan for this work were established. W.T.F. is grateful for the support of the Schmidt Science Fellows program, in partnership with the Rhodes Trust. We acknowledge Prof. G. Roth for fruitful discussions on results.